# Dark Matter in Nuclear Physics

George Fuller (UCSD), Andrew Hime (LANL), Reyco Henning (UNC),
Darin Kinion (LLNL), Spencer Klein (LBNL), Stefano Profumo (Caltech)
Michael Ramsey-Musolf (Caltech/UW Madison), Robert Stokstad (LBNL)

## Introduction

A transcendent accomplishment of nuclear physics and observational cosmology has been the definitive measurement of the baryon content of the universe. Big Bang Nucleosynthesis calculations, combined with measurements made with the largest new telescopes of the primordial deuterium abundance, have inferred the baryon density of the universe. This result has been confirmed by observations of the anisotropies in the Cosmic Microwave Background. The surprising upshot is that baryons account for only a small fraction of the observed mass and energy in the universe.

It is now firmly established that most of the mass-energy in the Universe is comprised of non-luminous and unknown forms. One component of this is the so-called "dark energy", believed to be responsible for the observed acceleration of the expansion rate of the universe. Another component appears to be composed of non-luminous material with non-relativistic kinematics at the current epoch. The relationship between the dark matter and the dark energy is unknown.

Several national studies and task-force reports have found that the identification of the mysterious "dark matter" is one of the most important pursuits in modern science. It is now clear that an explanation for this phenomenon will require some sort of new physics, likely involving a new particle or particles, and in any case involving physics beyond the Standard Model. A large number of dark matter studies, from theory to direct dark matter particle detection, involve nuclear physics and nuclear physicists.

One leading model for dark matter involves Weakly Interacting Massive Particles (WIMPs). These are theoretically well motivated and may be, for example, the lightest stable super-symmetric particle. WIMPs fit nicely into theories attempting to unify the fundamental forces of nature. The cornerstone of the attractiveness of the WIMP idea is that a massive particle with a weak interaction strength scattering cross section will freeze-out of thermal equilibrium in the early universe with about the right relic density to account for the observed dark matter.

However, the same weak interaction strength coupling that gives WIMPS the requisite relic density also means that there is a chance that these particles could be detected directly in the laboratory. We can estimate the rough number densities and fluxes of these WIMPS if they are indeed the dark matter in the Galaxy. The WIMPs could be detected as they recoil from nuclei in massive and ultra-pure detector targets operating deep beneath the Earth's surface. Alternately, WIMPS might be detected indirectly, for example, by the detection of ~ GeV energy neutrinos coming from WIMP annihilation in the center of the Earth, the center of the Sun, or the center of the Galaxy. Many of the experimental techniques and theoretical ideas underpinning WIMP models and detection schemes are rooted in nuclear physics.

There are other ideas about the nature of dark matter. These include axions and right-handed, "sterile" neutrinos. There are, likewise, vigorous experimental and observation searches underway for these dark matter particle candidates, as well as for WIMPs. And, as with WIMPs, nuclear physics figures prominently in both the theory and detection schemes for these particles.

Below we outline a number of frontier dark matter science issues, highlighting the key role of nuclear physics and nuclear physicists in the dark matter enterprise.

## Dark Matter & Baryogenesis in Nuclear Physics

The fundamental origin of baryonic and dark matter, and the connected "coincidence problem" given by the ratio of their average densities being of order unity, are major challenges for our understanding of the early universe. Almost forty years ago, Sakharov pointed out the three necessary microscopic conditions needed to generate, from symmetric initial conditions, a net baryon-antibaryon asymmetry [1]: the occurrence of interactions violating the conservation of the operators connected to (1) baryon number and (2) charge and charge-parity, and (3) a departure from local thermal equilibrium. A number of models have since been proposed, involving energy scales ranging from the electro-weak scale all the way up to the Planck scale.

Interestingly enough, the Standard Model (SM) itself encompasses all of the above mentioned "Sakharov conditions" at the time of the electro-weak phase transition in the early universe. It was recognized long ago, though, that the SM generates a much smaller baryon asymmetry in the universe (BAU) than is actually observed. The extension of the SM to TeV-scale super-symmetry, instead, was shown to be able to enhance the generated BAU, thanks to its extra, and largely unconstrained, particle content [3]. Supersymmetric electro-weak baryogenesis (EWB) in the Minimal Super-symmetric Extension of the Standard Model (MSSM) is a highly constrained – and therefore potentially very predictive – scenario: the lightest of the scalar partners of the top quark, for instance, must be lighter than its SM counterpart, and the Higgs boson mass must be very close to the current experimental lower limit.

While a connection between the origin of Dark Matter and that of ordinary matter is per se not necessary, models providing a common explanation certainly have a great appeal. In this respect, the MSSM also provides a very apt Dark Matter particle candidate, the lightest neutralino. Ref. [2] showed that the nature of the neutralino (e.g. its mass and composition) is of crucial importance to produce both the right amount of Dark Matter and to successfully generate the observed BAU. As detailed below, supersymmetric EWB, in its minimal formulation, leads to an ensemble of predictions, ranging from the permanent electric dipole moment of the electron, to neutralino DM detection rates, to searches for supersymmetric particles at colliders, that make the theory thoroughly testable from a variety of experimental standpoints.

Other scenarios where the generation of the BAU and Dark Matter are connected have been proposed in the literature, including Q-balls (topological soliton solutions to the MSSM field equations) in the context of Affleck-Dine baryogenesis [3], or exotic GeV-scale Dark Matter models featuring a non-vanishing baryon number [4]. These frameworks, however, do not typically lead to distinctive experimental signatures, and their theoretical appeal and motivation is weaker than those for supersymmetric EWB. Further extensions of the MSSM itself (in particular of its Higgs or gauge sector) provide further – and sometimes more natural, at least from the point of view of parameters fine-tuning – options; the drawback in this case is that the theory is way less predictive and constrained, and the range of possible experimental outcomes is not as unique as in the minimal setup. However, several instances exist where the phenomenology of next-to-MSSM scenarios, for the experimental outcomes we shall discuss next, are not dramatically different from the minimal version discussed here [5].

Figure1 illustrates, in the plane defined by the mass of the supersymmetric (fermionic) partner of the photons ($M_1$) and of the Higgs boson ($\mu$), the regions where both the relic neutralino abundance (green bands) and the amount of BAU (blue bands) correspond, respectively, to the observed amount of CDM and of baryon asymmetry [2]. The right panel shows the reach contours, in the same parameter space plane, for direct and indirect DM searches (specifically, the search for energetic neutrinos from the center of the Sun produced by pair annihilations of captured neutralinos, with the anticipated maximal Ice Cube sensitivity performance and the current constraints from the Super-Kamiokande data). It is worthwhile pointing out that neutrino telescopes are an extremely powerful probe of EWB in the MSSM: the available data from the SuperKamiokande experiment, in fact, already rule out a sizable portion of the parameter space, well beyond the anticipated phase-II reach of CDMS. In the foreseeable future, both ton-sized direct DM detectors and future neutrino telescopes will thoroughly probe the entire EWB parameter space (as shown in detail in Ref.[2]), even beyond the minimal MSSM paradigm.

In Fig. 2 we show the predictions for the permanent electron EDM, superimposed with the regions giving the correct BAU, for two different choices of supersymmetric parameters (in the right panel, $M_2$ indicates the mass of the fermionic superpartner of the Z). Even though one-loop EDM can be reduced to arbitrarily small values by taking the masses of the scalar fermions to be very large, the MSSM EWB scenario we consider here yields an *absolute lower bound* on the size of EDM induced at the two-loop level. Ref.[1] showed that loops involving fermionic superpartners and gauge and/or Higgs bosons generate sizable electron EDM's. In particular, maximal CP-violating phases are not compatible with BAU, as over the whole parameter space the resulting two-loop values for the electron EDM are larger than the current experimental limit. Smaller CP-violating phases, instead, can give rise to both the desired BAU and consistent values of the electron EDM, as shown in Fig.2. A scan over the whole space of parameters entailed the determination of an absolute lower bound on the value of the electron EDM, that, over the regions consistent with BAU, must be larger than $10^{-28}$ e cm. Future electron EDM can technically access such values, giving a second important handle on the supersymmetric EWB scenario. As consensus on the two-loop neutron EDM is lacking at present, we do not show here quantitative predictions for it. However, it is worth pointing out that in the somewhat analogue Split-Supersymmetry case, Romanino and Giudice [6] find that the ratio between the two-loop neutron and electron EDM lies in the range 2-10. We therefore expect a lower bound on the neutron EDM in the present context of the order of, or even larger than, $10^{-28}$ e cm.

Turning to collider searches, Ref. [2] showed that while the Tevatron reach will be very limited, the LHC can probe a sizable portion of the parameter space. A future International Linear Collider, with a center-of-mass energy as low as 500 GeV, will explore virtually all the viable parameter space of the model.

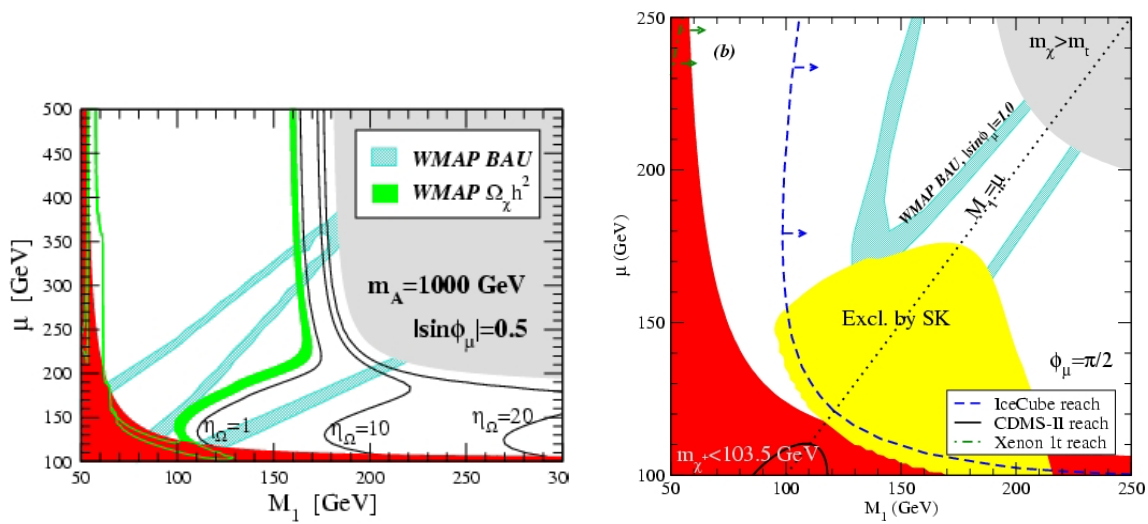

**Figure 1**: Left: the relic abundance of neutralino dark matter on the ($M_1$, $\mu$) plane. The red regions are excluded by the LEP2 chargino mass bound, while in the grey regions the scalar top becomes the LSP. The light green regions are where the neutralino thermal relic abundance falls within 2-$\sigma$ in the CDM abundance range determined by WMAP. The black curves are contours of constant , the number of "extra" neutralinos needed per thermally produced neutralino to bring low thermal relic abundance models into accord with the inferred CDM density. Right: Dark matter detection rates with maximal CP violation in the higgsino sector. The central shaded region is already ruled out by the SuperKamiokande data on the flux of energetic neutrinos from the Sun, while the regions lying to the right of the blue dashed lines will produce a sizable flux of neutrinos at IceCube. We also show the prospects for current (CDMS-II) and future, ton-sized (XENON-1t) direct detection experiments. The light blue shaded region corresponds to successful EWB.

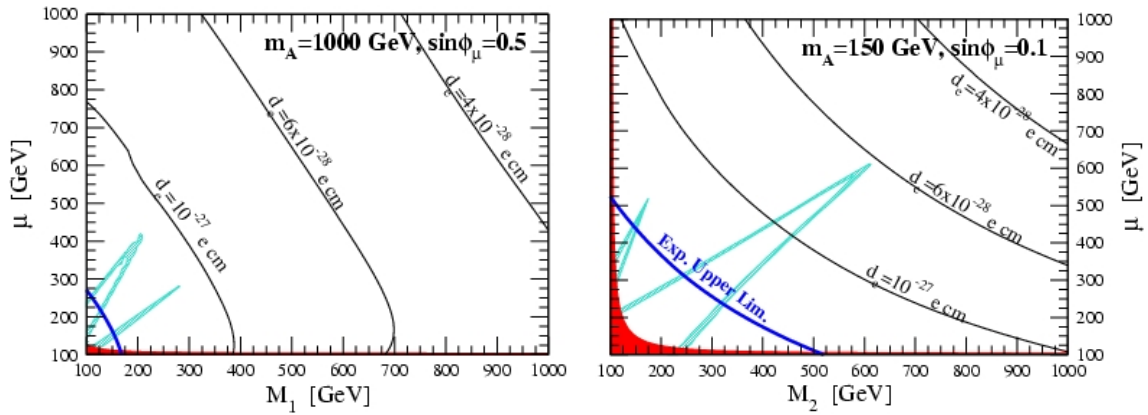

**Figure 2**: Iso-level curves of the electron electric dipole moment, on the ($M_{1,2}$, $\mu$) planes in the limit of heavy scalar fermions. The thick dark grey lines represent the current experimental upper limit. We also shade in light blue the 2-$\sigma$ regions corresponding to the correct BAU.

## Sterile Neutrino Dark Matter

Much has been learned in the last few years about the fundamental properties of neutrinos. (See the report on Neutrinos and Fundamental Symmetries.) Neutrinos have mass and their weak interaction (flavor) eigenstates are not coincident with their energy (mass) eigenstates. In fact, we have measured the mass-squared differences of the neutrinos and two of the three vacuum mixing angles in the unitary transformation between the energy and flavor bases. The remaining mixing angle is constrained and could possibly be measured in long baseline neutrino oscillation experiments or with a Galactic supernova neutrino signal. However, the CP-violating phase in this unitary transformation remains elusive.

The fact that neutrinos have mass is, in a sense, already physics beyond the Standard Model. It raises the possibility that there are right-handed neutrinos

(and left-handed antineutrinos). These are one way to construct "sterile" neutrinos which do not experience the ordinary weak interaction. We are at present completely ignorant of the properties of a putative right-handed neutrino sector. In particular we do not know their masses.

If these particles exist they may not be completely "sterile" on account of vacuum flavor mixing with ordinary "active" neutrinos (i.e., the electron, muon, and tau flavor neutrinos). The effective interaction strengths of these sterile neutrinos will be of order the product of this mixing and the standard weak interaction. It is likely, in fact, that these effective interaction strengths will be so weak that these particles might never come into thermal equilibrium in the early universe. This is why the rather stringent cosmological constraints on neutrino mass (which assume a thermal energy spectrum for the neutrinos and antineutrinos) may not apply to sterile neutrino states.

In fact, sterile neutrinos which have rest masses in the ~ 1 keV to ~100 keV range and which mix with ordinary active neutrinos at the level of one part in ~ $10^9$ to ~ $10^{12}$ in vacuum can have relic number densities which make them viable warm or cold dark matter candidates. These relic number densities are necessarily much smaller than those of the active neutrinos and photons.

Sterile neutrinos can be produced in the early universe through processes associated with active neutrino scattering-induced de-coherence. The production rate of sterile neutrinos in this case is essentially a product of the active neutrino scattering rate and the effective in-medium mixing fraction with active neutrinos. The scattering rate decreases (like ~ [temperature]$^5$) as the universe expands and cools, while the effective in-medium mixing angle for active-sterile neutrino flavor mixing goes down with increasing temperature. The result is that the sterile neutrino production rate is peaked. For mass and mixing parameters which give the correct relic dark matter densities, this peak occurs in or near the QCD epoch (temperature ~ 100 MeV). Moreover, the way in which the relativistic degrees of freedom at the QCD change with time/temperature are important ingredients in sterile neutrino relic density prediction. This is also a key issue in relativistic heavy ion research.

The best constraints on and probes of dark matter sterile neutrino rest mass and mixing properties come from X-ray astronomy. A sterile neutrino can decay by virtue of its mixing with active neutrinos in a non-GIM-suppressed process with a light active neutrino and a photon in the final state. The rate of this decay is proportional to 5 powers of the sterile neutrino mass. This process necessarily produces a photon with an energy that is half of the sterile neutrino rest mass, putting this photon squarely in the X-ray band.

Modern X-ray observatories such as Chandra and XMM Newton and the forthcoming Constellation-X have their greatest sensitivity for detection for photons with energies in the 1 to ~ 10 keV range, coincidently the range of most interest for sterile neutrino dark matter masses. Current observational bounds from the diffuse X-ray background and from X-rays from the Virgo cluster and Andromeda restrict the mass and mixing parameters of viable sterile neutrino dark matter to be in a regime where simple quantum de-coherence production with zero lepton number will not suffice.  However, if the universe has a lepton number significantly larger than the baryon number, then these experimental bounds would not apply.

The allowed sterile neutrino dark matter parameters imply interesting effects in core collapse supernovae and for the generation of pulsar kicks. These subjects are, in turn, active areas for research by the nuclear astrophysics community.

# Experimental Nuclear Physics & Dark Matter

Nuclear physics techniques are central to many of the experimental aspects of studies of dark matter.   This section will discuss some of the techniques of dark matter detection and discuss how nuclear physics is central to them.  These techniques naturally divide into two classes, direct and indirect detection.  Direct detection involves direct observation of dark matter interactions with nuclear matter in a laboratory experiment.  Indirect detection relies on observation of neutrinos from dark matter annihilation in the earth, sun, or other celestial bodies.

It is important to emphasize that direct and indirect detection are complementary; both have significant, but quite different model-dependencies.  The direct searches depend on the dark matter-nucleon cross-sections, while the indirect searches depend on secondary particle production in WIMP (or other particle) annihilation.

### Direct Detection

Many of the promising ideas for the direct detection of WIMP dark matter have been developed using techniques in experimental nuclear physics and include a strong synergy with other parts of the nuclear physics program such as the search for neutrinoless double beta decay and the detection of low-energy solar neutrinos.

Limits have been placed on the spin-independent cross-section for WIMP-nucleon scattering of about $10^{-43}$ cm$^2$, characteristic to that of a low-energy neutrino interaction. The most sensitive of these measurements has been made by the CDMS (Cryogenic Dark Matter Search) collaboration using an array of low-temperature, Ge-bolometers with a target exposure of ~ 50 kg-days. The challenge for future searches hinges on our ability to achieve sensitivities some three orders of magnitude beyond the existing state-of-the-art. This requires the development of novel detector technologies exploiting target masses of order 100 to 1000 kg [≡ 1 ton] and reaching unprecedented levels of radio-purity. To reach the desired sensitivity, "background-free" detectors must be created with the capability to identify a single event per ton per year.

Much of the challenge to realize sensitive dark matter detectors is in separating ubiquitous backgrounds (gamma rays and electrons) from the nuclear-recoil events characteristic of the WIMP signature. As the CDMS collaboration plan to scale in target mass in the SuperCDMS project, a suite of novel ideas has also emerged in recent years to exploit cryogenic liquids (LXe, LAr, and LNe) that hold great promise in their ability for discriminating signal from background and the relative ease for scaling to economically large target masses. Indeed, the U.S. plays a lead role in these developments through, for example, the XENON, LUX, DEAP, CLEAN, and WARP collaborations. Some of these detector concepts are also unique for their multi-purpose capability, for example the detection of neutrinos from the Sun and from core collapse Supernovae.

Investigators supported through the DOE Office of Nuclear Physics have also made many of the recent breakthroughs in the liquid-cryogen technology. This follows from a significantly large number of scientists in this country that have collaborated on low-energy neutrino physics experiments that require expertise in nuclear physics and ultra-low background detector technology in particular. In addition to the synergy with the experimental nuclear physics program, theoretical nuclear physics is also of import for the direct detection of WIMP dark matter. In particular, an interpretation of a limit or, indeed a signal, for WIMP dark matter in a given target requires knowledge of the nuclear form factor and quenching function characteristic of WIMP-nucleon scattering. It is prudent therefore to examine avenues to increase the support for direct dark matter detection across the various agencies and offices in order that investigators in this nation can continue to lead the world effort to identify the dark matter in the Universe.

### Indirect Detection

Indirect detection relies on the possibility that WIMPs may self-annihilate into stable standard model particles that may be detectable as anomalies in the cosmic-ray spectrum. In particular, WIMP-like dark matter has been studied in detail and the following potential signatures have been identified in cosmic-rays:
- Structure in the positron or antiproton spectrum in the 10-100 GeV range
- Observation of cosmic anti-deuterons or other antinuclei, above the tiny expected background
- Structure in the high energy gamma-ray spectrum (1-100GeV)
- High-energy neutrino fluxes from the sun and earth (discussed below).

The first two signatures are currently being searched for with significantly increased sensitivity by the PAMELA space-based experiment. They are expected to release their first results within a year. The AMS-02 experiment, currently under construction, will measure the same spectral features from the

International Space Station with even greater sensitivity. The GLAST satellite experiment and ground-based gamma-ray observatories such as VERITAS, MILAGRO, HESS, etc. will measure the gamma-ray flux with increased sensitivity in coming years. Indirect detection is an active field, but the signature that dominates is strongly model-dependent and it is crucial to pursue all these complementary measurements to ensure that all possible annihilation channels for WIMP dark matter are considered. The annihilation rate of WIMPs is also strongly dependent on the distribution of dark matter, scaling as the density squared. This makes indirect detection complementary to direct detection, since density perturbation can significantly enhance or suppress the flux of secondary cosmic-rays from WIMP annihilation.

### Neutrinos from the Sun

Nuclear Physics has a long history of studying neutrinos from the Sun, going back to the original Ray-Davis chlorine experiment.  More recently, the SNO experiment used solar neutrinos to confirm that neutrinos mix, and also measure (finally) the total neutrino production from in the sun, confirming the standard solar model.  This work has been extremely productive in both solar physics and neutrino physics.  Among other ways, this productivity was recognized when Ray Davis and Masatoshi Koshiba shared the 2002 Nobel Prize in Physics.

Neutrinos could also be produced by dark matter annihilation in the sun (or the earth).  Dark matter passing through the sun interacts weakly with it, and loses enough energy to be gravitationally captured.  Over time, it accumulates, and the density increases to the point where WIMP-WIMP interactions occur.  If WIMPs can self-annihilate, then in most models, these reactions produce neutrinos which can be observed in terrestrial detectors.  The neutrino energy spectra depend on the WIMP masses and decay modes, but, in most models, the neutrino energies are in the 10 GeV to 10 TeV range.

Searching for this higher energy neutrino production in the sun is a natural conceptual extension of these efforts.  Both the Super-K and AMANDA experiments have searched for these higher energy neutrinos (Ackermann, 2006).  MILAGRO has searched for gamma-rays from WIMP annihilation near the sun (Atkins, 2004). Using similar water/ice Cherenkov techniques, experiments like IceCube (Ahrens, 2001), the near-future Mediterranean experiments (NESTOR, ANTARES, and NEMO) and the European $Km^3$ initiative will extend the solar neutrino energy spectrum from the MeV range to the GeV or TeV range.  Although the neutrino production mechanisms are somewhat different, all of these experiments probe the reactions taking place in the sun, and measure the energy flow due to neutrinos.

### Neutrinos from the Earth

The possibility of observing neutrinos from the earth goes back at least 50 years, when geologists and physicists discussed observing neutrinos from uranium and thorium decay. Last year, Kamland finally observed these neutrinos. This observation opens the door for more detailed measurement of both the absolute level and the distribution of nuclear activity in the Earth.

Similarly, searches for neutrino annihilation from WIMP annihilation in the earth's core are natural extensions of this effort, looking at neutrino production in a different energy range, again using water Cherenkov technology. AMANDA has searched for WIMP annihilation from the earth; IceCube and the European initiatives can extend these limits by an order of magnitude.

### Axion Detection

Axions are another excellent example of the importance of nuclear physics in determining the nature of the dark matter. Standard Model QCD predicts that the strong interactions should violate the discrete CP symmetry, resulting in a substantial CP-violating neutron EDM. Attempts to measure the neutron EDM, however, constrain this CP-violation to be extremely small. The most elegant solution to this so-called Strong-CP Problem is the suggestion by Peccei and Quinn [11] that there is a new global U(1) symmetry in the Standard Model that is spontaneously broken at the unknown energy scale $f_{PQ}$. With the addition of this spontaneously broken symmetry, it can be shown that the expected CP violation is naturally removed. As pointed out by Weinberg and Wilczek [12], a consequence of this SSB is a pseudo-Goldstone boson that was named the axion.

Since the mass of the axion is unspecified by theory, a number of searches have been carried out over many decades of energy. The strongest astrophysical and cosmological constraints currently limit the mass to be in the range of 1 $\mu$eV to 10 meV. Axions in the lower end of this mass range would not only explain CP conservation in the strong interactions, they would also be a significant fraction of the galactic dark matter.

So far, only microwave cavity axion searches would be sensitive to $\mu$eV mass axions. The Axion Dark Matter eXperiment (ADMX) is currently working to search the lowest two decades of allowed mass with sensitivity to even the most pessimistic axion models [13]. The narrow-bandwidth nature of this search means that it will require several years to scan this region. Other experiments, such as the CERN Axion Solar Telescope (CAST) [14] and PVLAS [15], are broadband, but lack the sensitivity of the microwave cavity searches.

Nuclear physics is crucial to constraining the mass of the axion. Detailed understanding of the nuclear processes that take place in stellar interiors and during supernovae explosions places fairly stringent limits on the production of new light particles such as axions. Further improvements will be important to future axion searches.  The CAST experiment, which searches for axions produced in the sun, obviously depends heavily on models of nuclear processes in the solar interior.  Microwave cavity searches will also benefit; given the slow nature of these searches, even modest improvements of the mass constraints can cut down the required scan time substantially.  At the same time, axion detection is fundamentally important to Nuclear Physics because it would explain the long-standing mystery of CP-conservation in the strong interactions.

## Nuclear Physics Opportunities with Dark Matter

Direct and indirect searches for dark matter complement each other.  In the absence of clear experimental or theoretical guidance, dark matter could take many, many forms.  So, a broad-based approach has the best chance of success.  In this respect, both direct and indirect searches are important, since they probe different types of dark matter (with some overlap, of course).  Direct searches for WIMPs and axions are in progress, while indirect searches look for gravitational capture and subsequent annihilation of any type of massive particles.   For WIMP masses above about 100 GeV, IceCube will be more than an order of magnitude more sensitive than existing direct detection experiments.

### Indirect Searches for WIMP Annihilation in the Earth and Sun

Construction of the IceCube neutrino observatory will open up a new window of sensitivity in searches for WIMP annihilation in the earth and the sun.  Experiments like SuperKamiokande and AMANDA have set limits on the possible flux of neutrinos coming from the earth and the sun.  However, these limits have been limited by the relatively small size of these detectors.  IceCube is more than an order of magnitude larger than AMANDA.  It also has much better directional accuracy, and so will have lower backgrounds than AMANDA; these factors will allow it to set correspondingly better limits, particularly for WIMPS from the sun.  This has a particular attraction for searches for heavier WIMPs, which are better captured by the sun.  Figure 3 compares the limits achieved with existing experiments, and the limits expected from IceCube and ANTARES (one of the Mediterranean experiments).

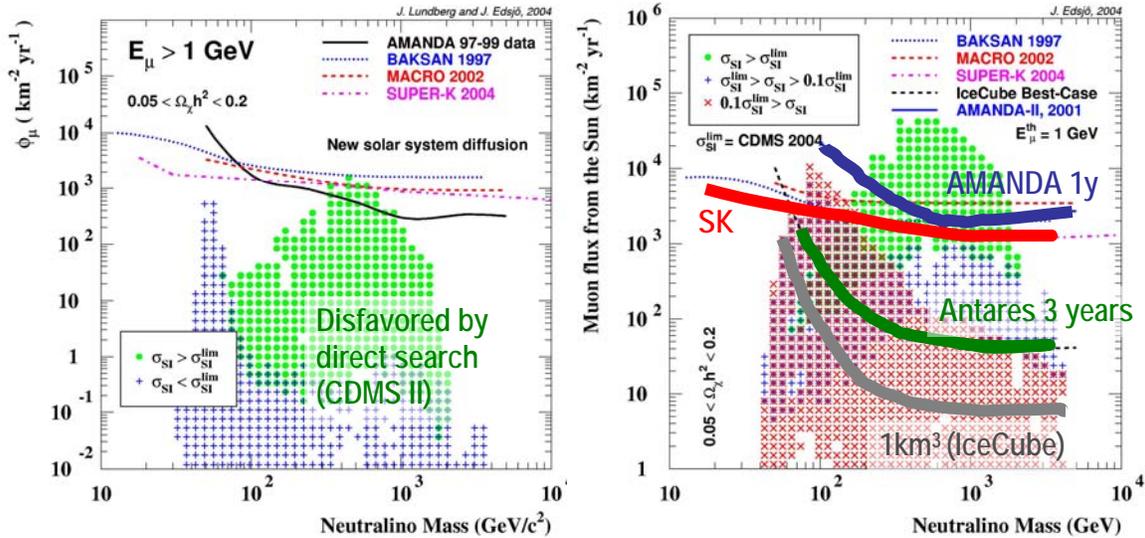

Figure 3. The current and expected WIMP search limits for underwater/inice detectors for neutrinos from the earth (left) or the sun (right). The x-axis here is the neutralino mass, while the y axis is the normalized muon flux. The points show the predictions in the minimal supersymmetric models; the colors indicate the cross sections for WIMP-nucleon interactions (and thus their potential to be discovered in direct experiments). Green is disfavored by experimental searches; blue covers cross sections from just below the experimental limit to 10% of the experimental limit, and red shows smaller cross sections. The two plots are quite different because the sun and the earth have different gravitational potentials, and because the experimental challenges of finding neutrinos from the center of the earth and the sun are different.

## Recommendations & Conclusions

Dark Matter is one of the most compelling physics problems of the last two decades. The form that dark matter takes has important implications for nuclear physics, particle physics, astrophysics, and cosmology. Similarly, techniques from all of these areas are being brought to bear on the problem, and a combined effort is likely needed to finally pin down the nature of dark matter.

This document has focused on the nuclear physics aspects of dark matter. Nuclear Physics offers a variety of techniques to probe dark matter, and many experimental studies are now being pursued. The two broad classes, direct and indirect searches are complementary. In recent years, many novel ideas for direct detection have emerged; they hold promise to increase our sensitivity by several orders of magnitude, spanning the parameter space predicted by the MSSM. Similarly indirect detection studies at the next generation neutrino telescopes offer the promise of orders-of-magnitude increase in sensitivity, again covering the MSSM parameter space. These detectors are already being built; the additional effort required for dark matter searches is modest.

*We recommend that these efforts in direct and indirect detection should be enthusiastically supported by the nuclear physics community, and that funding be made available for nuclear physicists to work in these areas.*